# Culture and constitutional compliance

Jerg Gutmann,[*] Anna Lewczuk-Czerwińska,[†]

Jacek Lewkowicz,[‡] and Stefan Voigt[§]

[*] Institute of Law & Economics, University of Hamburg, Alsterterrasse 1, 20354 Hamburg and CESifo, e-mail: jerg.gutmann@uni-hamburg.de.

[†] Faculty of Economic Sciences, University of Warsaw, Długa 44/50, 00-241 Warsaw, e-mail: alewczuk@wne.uw.edu.pl.

[‡] Faculty of Economic Sciences, University of Warsaw, Długa 44/50, 00-241 Warsaw, e-mail: jaceklewkowicz@uw.edu.pl.

[§] Institute of Law & Economics, University of Hamburg, Alsterterrasse 1, 20354 Hamburg and CESifo, e-mail: stefan.voigt@uni-hamburg.de.
This research is part of a Beethoven project funded by the Polish National Science Centre (NCN, #2016/23/G/HS4/04371) and the Deutsche Forschungsgemeinschaft (DFG, #381589259). The authors thank Abishek Choutagunta, Marek Endrich, Niclas Berggren, Katharina Luckner, Hashem Nabas, Eva Nissioti, Dina Rabie, Eman Rashwan, Roee Sarel, Betül Simsek, and participants of the Association for the Study of Religion, Economics, and Culture Conference, the German Law and Economics Association Conference, the European Public Choice Society Conference, NOUS, and research seminar participants in Hamburg, Milan, and Paris for helpful comments and suggestions.

# Culture and constitutional compliance

**Abstract**

Constitutions as the formal foundation of a country's legal and political system have important economic and political effects. Yet, we still know little about why constitutions set effective constraints on politicians in some societies, while being largely disregarded in others. Here, we ask if national culture matters for constitutional compliance. We study a cross-section of 115 countries, making use of novel indicators of constitutional compliance. We find that societies with a more individualistic population exhibit higher levels of compliance. These results are robust and extend to instrumental variables estimations. They imply a novel transmission channel from cultural traits to long-term economic development: individualistic national culture increases the credibility of constitutional self-commitments. Our analysis also supports the more general idea that the effects of formal institutions depend on the informal institutional environment in which they are embedded. Regarding religion, our results are consistent with past research that attributes the lack of development in the modern Muslim world to deficient institutional quality.

## 1. Introduction

Reading a country's constitution to learn about its political and legal system is informative in some cases and unenlightening in many others. But why do some governments behave strictly in accordance with the rules laid down in their country's constitution, while others do not? Although this question concerns the basic functioning of countries' political and legal systems and is, therefore, of great policy relevance, we still know surprisingly little about the causes of constitutional compliance.

Economists have for some time pointed toward the effectiveness of government constraints as a key challenge in economic development. Acemoglu and Robinson (2019, 2023) call states that are strong enough to enforce laws while simultaneously being restrained in their actions "shackled Leviathans". They represent a desirable but fragile equilibrium between too much state capacity and too much capacity of society. Acemoglu and Robinson (2022) argue that a culture that is conducive to sustaining a shackled Leviathan should have three traits: (1) a suspicious attitude towards those who are politically powerful, (2) trust in impersonal institutions and the political hierarchy that is associated with them, and (3) a strong belief in popular sovereignty. Here, we are interested in empirically evaluating which national cultural traits are conducive to putting the Leviathan in constitutional shackles. While a shackled Leviathan could, in theory, exist without an effectively enforced constitution, modern societies predominantly rely on formal law to organize their social and political order and to set limits on legitimate government action (Elkins et al. 2009; North and Weingast 1989). In line with this reasoning, empirical evidence confirms that countries in which the government respects the constitution enjoy higher growth rates (Lewczuk and Metelska-Szaniawska 2025).

For a long time, culture has been ignored in mainstream economics. This has radically changed in recent years (see, e.g., Acemoglu and Robinson 2025; Alesina and Giuliano 2015; Giuliano 2020; Guiso et al. 2006; Kuran 2018). The academic quest for the deep determinants of economic development has certainly contributed to economists' newfound interest in the economic and social consequences of culture. Tabellini (2008a, 2008b, 2010) conducted some of the seminal studies on how culture, institutions, and economic development are linked to each other. He shows that having a favorable national culture is associated with better legal enforcement and higher income. Recent research by Gorodnichenko and Roland (2011, 2017, 2021) is also closely related to our research question. They find that one particular cultural trait, namely individualism, not only explains cross-country differences in growth and productivity, but also in innovativeness and the establishment of democracy.

We add to the extant literature on culture, institutions, and development by focusing on the relationship between culture and a very specific group of institutions, namely those codified in constitutions (Elkins and Ginsburg 2021). Constitutions frequently include the most important institutions used to organize a society's economic, political, and social life. If institutions are defined as commonly known rules for structuring recurrent interactions that are endowed with a sanctioning mechanism, then constitutions can be thought of as the most fundamental layer of a society's codified institutions. Thus, our research question deals with an under-researched aspect of the relationship between culture and formal institutions (see Alesina and Giuliano 2015 for a survey of that literature). More specifically, we are interested in the cultural traits of societies that determine whether a government is likely to comply with the promises made in the constitution. In terms of constitutional promises, we focus on protections granted to citizens (e.g., regarding property ownership, physical integrity, free political association, or free expression), particularly vis-à-vis their government.

While a literature has emerged in recent years that has sought to explain differences in constitutional compliance across countries (e.g., Bologna Pavlik and Young 2026; Choutagunta et al. 2024; Grajzl et al. 2025; Gutmann et al. 2024a; Kantorowicz and Voigt 2025; Lewkowicz and Lewczuk 2023; Lewkowicz et al. 2024), these studies have not analyzed its possible deep (or long-run) determinants—most of them exclude such factors explicitly from their analysis by controlling for country fixed effects. In other words, they have focused on factors that can change from year to year, whereas we are interested here in the relevance of fundamental cultural traits, which typically vary little, even over the course of decades (Schulz et al. 2019; Williamson 2000). This is, therefore, the first study to ask whether factors outside political decision makers' control shape the extent to which governments comply with their country's national constitution.

Our sample covers up to 115 countries for which data on both culture and constitutional compliance are available. To measure national culture, we rely on data sources that are widely recognized in the economics literature. "Generalized morality" is an indicator capturing values relevant to economic transactions and was introduced by Tabellini (2008a, 2010). Individualism, power distance, and long-term orientation have been measured by Hofstede et al. (2010) and are the most well-established indicators of culture in cross-national economic studies. Finally, Muslim population shares are used as a proxy for the influence of the most important major religion that has been linked to systematic cross-country differences in institutional quality and political accountability (Kuran 2010, 2018). To measure constitutional compliance, we rely on an indicator developed by Gutmann et al. (2024b). It captures the degree to which the provisions in a national constitution are implemented in practice. The 14 types of constitutional rules underlying the indicator cover four distinct legal areas (property rights and the rule of law, political rights, civil liberties, and basic human rights). To address endogeneity concerns that may arise even though the fundamental cultural traits we are interested in hardly vary over time, we estimate instrumental variables regression models with instrumental variables that have been validated in past studies.

We find that having an individualistic population is conducive to higher constitutional compliance by the government, whereas a higher Muslim population share is linked to reduced constitutional compliance. These results are robust to different estimators and model specifications. Both effects are even robust to controlling for democracy, although that reduces the effect size by almost 50%. Generalized morality, power distance, and long-term orientation exhibit the expected sign and are significant in individual estimations, but their estimated effects are not robust. Among other major religions, only Protestantism also seems to be associated (positively) with constitutional compliance. That would be consistent with claims about Protestant doctrine encouraging independent thought and religious individualism.

The remainder of this article is structured as follows. In the next section, we define our key terms and discuss general theoretical connections between culture and constitutional compliance, before hypothesizing which cultural traits are particularly conducive or detrimental to constitutional compliance. Section 3 describes our data and the estimation approach. In Section 4, we discuss our empirical results, and Section 5 concludes.

## 2. Concepts and Theory

### 2.1. Definitions

Culture has been defined in countless ways. Here, we follow Guiso et al. (2006) in defining culture as "those customary beliefs and values that ethnic, religious and social groups transmit fairly unchanged from generation to generation" (Guiso et al. 2006, p. 23). Two aspects of this definition are worth emphasizing. First, it includes both beliefs and values. Whereas beliefs describe people's understanding of how the world works, values refer to their convictions concerning how it ought to be organized. The second aspect is that culture is largely time-invariant (or slow-moving, as described by Roland 2004). Williamson (2000) proposes the very crude rule of thumb that culture and informal institutions change once in a century, whereas formal institutions, such as constitutions, change about once in a decade. This feature of culture is important to consider when studying its potential influence on institutions or on economic development. In line with this definition, many empirical studies treat culture as plausibly exogenous. However, there are also some established instrumental variables, based on geography or major historical events, that can make causal identification even more credible.

Constitutions define constitutional actors and allocate competences and obligations to them. In modern times, constitutions typically also contain a catalog of rights that the country's citizens or inhabitants are supposed to enjoy. We speak of constitutional compliance, if the members of the executive branch of government follow and enforce the rules laid down in the constitution. To ascertain the level of constitutional compliance, one has to compare the *de jure* provisions of a constitution with the constitutional reality, as implemented by the executive (see, e.g., Voigt 2021).

Two aspects of our definition of constitutional compliance are noteworthy: First, we do not resort to some universal rights standard as a benchmark, but we consider only the rules in the national constitution of each country. This implies that every country defines its own benchmark. While data constraints limit our measurement of constitutional compliance to 14 types of constitutional rules, a government is judged only for its compliance with those rules that are established in its national constitution. Second, constitutional compliance is related to but not synonymous with the rule of law. The central trait of the rule of law is that all members of society are subject to the same general and predictable rules. If a government complies with the national constitution, its actions do not automatically follow the rule of law, and vice versa.

Although constitutional compliance, as operationalized in this study, leaves the decision which rights and principles shall be protected to every nation, its measurement is still partially based on a normative judgement—in the sense that the list of rights and principles under consideration is comprised of important elements of a Western conception of limited government. A government that is, for example, in full compliance with a totalitarian constitution or with religious scripture that enjoys constitutional status might still not reach a high constitutional compliance score, because these rules might not be among the 14 rules on which the constitutional compliance indicator is based.[1] While this is a conceptual limitation, in practice, countries all around the world consider the right to life, the prohibition of slavery, or freedom of movement to be fundamental rights that deserve to be constitutionally protected.

---

[1] While highlighting this information is important for the correct interpretation of our results, it is by no means accidental that constitutional compliance is not based on a more nihilistic operationalization. If the strict enforcement of a totalitarian constitution that prescribes, e.g., a fascist political and social order would be considered constitutional compliance, then the expected benefits of an enhanced credibility of the government would be largely irrelevant (see Bernholz 1991).

### 2.2. Culture and Constitutional Compliance

Although constitutions are the most basic layer of the formal institutions of society, their implementation is not guaranteed, as there is no formal authority that could ensure the proper implementation of constitutional rules. Constitutions must be self-enforcing, i.e., members of the executive cannot make themselves better off by overstepping constitutional constraints (Mittal and Weingast 2013; Weingast 1997, 2005). Compliance with constitutional rules, thus, critically depends on politicians' expectation that others are willing and able to incur costs to punish transgressions against the constitution (Gutmann et al. 2025; Hurwicz 2008; Weingast 1997, 2005). This trait of constitutional rules makes culture a potentially important factor in explaining constitutional compliance.[2]

Even if members of the executive consider reneging on constitutional rules, constitutional compliance can still be secured via four mechanisms: violations can be sanctioned by (1) the citizens, (2) veto players, (3) the politicians themselves, and (4) the international community. Any behavior that increases the cost of noncompliance from the point of view of the executive can serve as a sanction. Our analysis focuses on the first three of those mechanisms, as they can be influenced by national culture. More formally, the political leader's utility function can be summarized here as $U = U(S, C, M)$, where $S$ is the political support cost due to citizen opposition, $C$ is the political capital cost due to veto player opposition, and $M$ is the morality cost imposed by the politician's own ethics. $U$ is declining if policies that violate the constitution cause costs in the form of $S$, $C$, or $M$. Next, we explain how these costs may depend on the dominant culture that is characteristic of a nation's citizens and politicians.

[2] See also our discussion of Acemoglu and Robinson's "narrow corridor" in the introduction.

Citizens can react to constitutional noncompliance by politicians in a variety of ways. They can, for example, spontaneously participate in demonstrations or strikes or they can organize themselves in more permanent civil society organizations (see, e.g., Gutmann et al. 2021). To organize protest that is costly to the members of the executive (imposing the cost $S$ in terms of lost political support), citizens need to coordinate their behavior. The ability to overcome the dilemma of collective action might be aided by specific cultural norms, which, for example, set focal points (Olson 1965; Schelling 1960; Weingast 1997). Solidarity norms prescribing that one should come to support those who have been treated wrongly, in particular by representatives of the state, are but one example of such norms (see also Guiso et al. 2011 on civic capital). Cultural norms may not only facilitate protest, but they can also affect the propensity of citizens to vote, to participate in civil society organizations, as well as their preferences as voters. At least in democracies, citizens can punish politicians or their parties by voting them out of office (Fiorina 1978). All these potential actions by citizens can set incentives for politicians to comply with the constitution. This is the primary idea of Acemoglu and Robinson (2019, 2022) for how the Leviathan can be put in shackles, as discussed in the introduction.

The second mechanism is related to the behavior of veto players. Veto players are actors who have the capacity to block a change to the political *status quo* (Tsebelis 2002). Prominent examples of veto players are courts, regulatory agencies, or regional governments. It is important to note that these veto players, to be effective, need to be independent, both from the government and from powerful interest groups. Applied to constitutional compliance, this idea could refer to the ability of the legislature or the judiciary to stop members of the executive from violating constitutional rules. If cultural norms demand or encourage such interventions from the members of these branches of government and they have the capacity to do so, veto players can be expected to impose a cost of $C$ in terms of political capital needed to overcome veto player opposition and this expectation incentivizes constitutional compliance by the executive.

It is not necessarily only constraints imposed by others that make politicians follow constitutional prescriptions. If their ethics dictate strict compliance with rules, then breaking a rule to make one better off might be prohibitively costly in many cases in terms of moral or psychological costs $M$. This is the third mechanism by which culture can change the likelihood of constitutional compliance (Congleton 2020 analyzes the relevance of ethics for politicians' behavior).

Although these three mechanisms can be clearly delineated from each other in theory, it is difficult to ascertain their individual relevance empirically. Rather than probing into such transmission channels, we ask next which cultural traits are likely to enhance or hinder compliance with the constitution by members of a country's executive. However, in an extension of our empirical analysis, we will come back to the described mechanism and we will test whether different cultural traits influence constitutional compliance via their effect on civil society or on political veto players.

### 2.3. Cultural Norms Conducive to Constitutional Compliance

Today, most constitutions claim that everybody is to be treated alike, independent of their personal characteristics, such as age, gender, race, religion, or income (Elkins et al.

2009, p. 86). Cultural norms congruent with such stipulations may be expected to lead to higher constitutional compliance (see, e.g., Berkowitz et al. 2003a, 2003b on the effectiveness of legal transplants). However, we are not interested in measuring the match between cultural norms and constitutional rules (as, e.g., stressed by Acemoglu and Jackson 2017). Instead, we try to identify cultural norms that are conducive (or detrimental) to constitutional compliance largely independent of the constitutional rule in question. We argue that three common national cultural traits can be expected to play an important role: (1) the level of differentiation between ingroups and outgroups in society, (2) the rate at which future benefits of today's compliance are discounted, and (3) whether a dominant religion claims supremacy over humanly devised law and demands state enforcement of inequalities among members of society.

Intergroup differentiation

Individualism describes a preference for a loosely knit social framework in which individuals are expected to take care of themselves and their immediate family. In contrast, collectivism describes a preference for a tightly knit social framework in which individuals expect their relatives or members of their in-group to look after them in exchange for loyalty to that family or social group (Hofstede et al. 2010). Individualistic societies are characterized by moral values that reflect universal impersonal principles, such as fairness, individual rights, and justice. These principles emphasize the welfare of all individuals in society equally. Collectivistic societies exhibit a morality that is based on communal values, such as in-group loyalty or the moral relevance of betrayal and respect tied to particular identities, groups, and relationships (see Enke 2019). We understand the individualism-collectivism distinction as variation along a continuous analytical dimension rather than as a rigid binary taxonomy of societies. Such an approach is consistent with recent research in economics that treats culture, moral systems, and individualism as measurable traits (Ang 2019; Bazzi et al. 2020; Enke 2019; Ezcurra 2021; Kristjánsdóttir et al. 2017, Rawal and Dash 2026).[3]

Citizens in more individualistic societies are, thus, more likely to expect constitutional compliance from their political representatives.[4] In the language of Acemoglu and Robinson (2022), they trust impersonal institutions and political hierarchies built on them. Accordingly, individualistic societies should be more prone to entrust independent veto players with power. In contrast, keeping up tradition plays an important role in collectivistic cultures with their emphasis on time-honored community values. A constitution that attempts to legitimize the state with rational thought might even be met with indifference or be outright rejected by large parts of collectivistic societies, if the constitution's content does not match society's often discriminatory informal norms. Also, the emergence of a vibrant and open civil society is less likely in a collectivist society,

---

[3] Hofstede's approach has been criticized from various angels with titles such as "Hofstede's imagined cultures" (McSweeney 2024) or "Hofstede never studied culture" (Baskerville 2003). The two most frequently criticized aspects refer to the non-appropriate equation of culture with nation and the measurement from an outsider's perspective – instead of deriving concepts from within a culture, as preferred by anthropologists. Yet, Hofstede's original 1980 book is among the 25 most cited books in the social sciences (Green 2016). Smith et al. (2006, 33ff.) offer a very balanced evaluation of Hofstede's project. They point out that some of the critiques do not consider that Hofstede was very careful in linking individual level data to the national level, and some of the precautions he took (e.g., by taking acquiescent response bias into account) are not sufficiently appreciated by his critics. Smith et al. (ibid., 46) also point out that Hofstede's individualism measure and similar measures, such as autonomy by Schwartz or self-expression by Inglehart, are highly correlated (r=0.64 and r=0.74).

[4] Individualistic societies not only demand more constitutional compliance, but their governments are also be better able to enforce the constitution, due to higher levels of state capacity (Wiśniewska-Kuźma 2026).

as social organization rather follows existing, identity-based social structures that are not open to everyone. Therefore, we expect that:

**H1:** Constitutional compliance is higher in more individualistic societies.

Cultures emphasizing that different people ought to have different roles in life and especially cultures endorsing a strict social hierarchy among people are likely incompatible with both equal rights guaranteed by a constitution and constitutional constraints imposed on the political class. This can be a major reason for constitutional noncompliance, especially if forms of equal treatment are explicitly stipulated in the constitution. As explained above, such rules are typically part of modern constitutions.

Power distance expresses the degree to which the less powerful members of a society accept and expect that power is distributed unequally. It implies the acceptance of established rules and a hierarchical order in which everyone has their place without further justification. In contrast, members of societies with low power distance strive to equalize the distribution of power and demand justification for any remaining inequalities in power (Hofstede et al. 2010). According to Acemoglu and Robinson (2022) that skepticism is important for the ability to shackle the Leviathan. The unconditional acceptance of hierarchical orders gives politicians leeway in dealing with the constitution, whereas citizens and civil society organizations in societies with low power distance would challenge politicians for such transgressions. Thus:

**H2:** Societies characterized by higher levels of power distance exhibit lower levels of constitutional compliance.

Van de Vliert (2020) describes collectivism and power distance as two manifestations of the same latent cultural trait: intergroup differentiation—the extent to which people are treated as members of either ingroups (same family, clan, community, tribe, social class, nation, …) or outgroups, rather than as individuals. More intergroup differentiation implies sharper us vs. them-boundaries, larger psychosocial distance and greater behavioral differences between familiars and strangers (collectivism) and between higher-ups and lower-downs (power distance). We can, thus, summarize the argument behind our first two hypotheses as low intergroup differentiation being conducive to the acceptance and appreciation of a system of universal (constitutional) rules that governs interpersonal relationships in an impartial fashion. However, that does not imply that one of these dimensions cannot be more important than the other for constitutional compliance.

Long-term orientation

The second type of cultural trait that we argue is relevant for constitutional compliance is impatience, i.e., a high time discount rate, as those not complying might want to reach a certain goal immediately and irrespective of its longer-term opportunity costs. Falk et al. (2018) show that patience is the single most important standard preference in explaining economic development. Doepke and Zilibotti (2008) even use the term patience capital to underscore the relevance of patience for productivity (see also Galor and Özak 2016). Recent evidence by Sunde et al. (2022) points in the same direction, linking patience to physical and human capital accumulation, productivity, and long-term development. A similar intertemporal logic applies to constitutional compliance, as enforcing constitutional constraints may obstruct immediate political goals, but it preserves long-term benefits such as institutional stability, credible commitment, and predictable limits on government power. This interpretation accords with models in which veto players help voters limit abuse of power (Persson et al. 1997), while electoral accountability alone may otherwise induce pandering to current public opinion (Maskin and Tirole 2004). We argue that if a culture attributes high value to patience, this should favor constitutional compliance, as legal certainty created today generates benefits for many years to come. Complying with constitutional rules today frequently entails opportunity costs in the present that can only be outweighed by future benefits if these are not discounted too much by the decisionmaker. Following Hofstede et al. (2010), we refer to this national cultural trait as long-term orientation and hypothesize that:

**H3:** More long-term oriented societies exhibit more constitutional compliance.

Religion

Religions are important in transmitting beliefs and values from generation to generation and are, therefore, often treated as a form of culture. Two traits of religions seem particularly relevant here. The first is that some religions prescribe that not all members of society should enjoy the rights typically found in modern constitutions. Apostates, members of other religions, or women might be subject to religiously motivated legal discrimination. The second trait depends on how a religion conceptualizes the relationship between individuals and the state. If a religion, for example, endorses obedience to rulers, even if they do not comply with the constitution, then constitutional compliance would be seriously undermined. That is consistent with Acemoglu and Robinson's (2022) emphasis on the importance of a strong belief in popular sovereignty for constraining the Leviathan.

Islamic constitutions frequently declare the supreme constitutional status of Islamic law, which demands noncompliance with parts of the constitution that appear to contradict Islamic legal principles (e.g., Bernholz 1991; Gouda and Gutmann 2021). Modern Islamic constitutions frequently incorporate "Islamic review clauses" or "Sharia guarantee clauses", which mandate that no state law can be repugnant to the principles of Islam (Ahmed and Ginsburg 2014; Lombardi 2013). However, the interpretation of what the divine law prescribes is ambiguous and critically depends on who interprets the clauses (Lombardi 2013; Quraishi-Landes 2015; Brown and Revkin 2015). Islamic review is a tool used by state actors to override standard constitutional protections, weakening overall constitutional compliance (Lombardi 2013). Ahmed and Ginsburg (2014) demonstrate that constitutions with Islamic review clauses include more rights as compared to Muslim constitutions without such clauses. This may contribute to lower levels of constitutional compliance: When a constitution promises a variety of liberal rights while simultaneously establishing complex mechanisms of religious supremacy, the state faces conflicting mandates, which may lead to selective enforcement.

Among the major contemporary world religions, Islam is the one that is best-known for prescribing inequalities in legal status (Gouda and Potrafke 2016; Gutmann and Voigt 2015, 2018a; Powell et al. 2021). Some interpretations of Islamic law sanctify a number of inequalities related to, inter alia, unequal treatment of men and women, Muslims and non-Muslims, and other social groups, which conflicts with the rule-of-law principle of equal treatment (Gutmann and Voigt 2015, 2018a). Therefore, constitutional compliance with respect to equality, rights, or judicial protection is likely to be weaker in Muslim states. Accordingly, we hypothesize that:

**H4:** Societies strongly influenced by Islam exhibit less constitutional compliance.

## 3. Empirical Analysis

### 3.1 Data and Empirical Design

In our empirical analysis, we estimate cross-sectional models of the form:

$$Compliance_i = \alpha + \beta \times Culture_i + \gamma \times X_i + \varepsilon_i \qquad (1)$$

where $Compliance$ is an indicator of the government's average constitutional compliance in country $i$ between 2010 and 2019. $Culture$ is an indicator of the extent to which a particular cultural trait is prevalent in that country. The vector $X$ comprises control variables that are plausibly related to both culture and constitutional compliance.

Our dependent variables measure the extent to which a government complies with its national constitution. We rely on the novel Comparative Constitutional Compliance Database (CCCD, see Gutmann et al. 2024b), which combines information from two publicly available sources. The Comparative Constitutions Project (CCP) records *de jure* constitutional rules and the Varieties of Democracy (V-Dem) project measures their *de facto* implementation. Information from CCP is used to establish whether a country's constitution includes a specific rule, such as prescribing an independent judiciary or the prohibition of torture. Variables from V-Dem are then matched to these dummy variables to ascertain if the rules that are part of the constitution are complied with in practice. Scores are strictly monotonically increasing (decreasing) in the number of complied with (violated) constitutional rules, implying that they are highest when all constitutional rules are complied with and lowest if none of them are.

Dependent variable

The CCCD measures compliance in four legal areas based on 14 types of *de jure* constitutional rules and their *de facto* implementation. Principal factors are calculated for (1) property rights and the rule of law, (2) political rights, (3) civil liberties, and (4) basic human rights. As the calculated factor loadings are very similar within each dimension, compliance scores are virtually identical if, instead of applying factor analysis, the mean value is calculated. These four indicators are then aggregated by taking the mean value into one overall indicator of constitutional compliance, which serves as the main dependent variable in our empirical analysis. Version 2.0 of the CCCD covers the period between 1900 and 2020. As we are interested in the long-run cultural determinants of constitutional compliance, we eliminate noise in the form of short-run fluctuations in the data by taking the mean value for the period 2010 to 2019, i.e., the period between the Great Recession and the COVID-19 pandemic.

Cultural dimensions

To measure the prevalence of national cultural traits, we rely on data by Geert Hofstede, who initially surveyed IBM employees in the 1960s and over time extended the indicators and data sources (see Hofstede et al. 2010). Although Hofstede's original data were collected to measure corporate culture, they have since been validated in numerous studies as a measure of national culture. The indicators have been used extensively in the social psychology literature and are increasingly used in economic research (see, e.g., Debski et al. 2018; Falk et al. 2018; Galor and Özak 2016; Gorodnichenko and Roland 2011, 2017, 2021; Gründler and Köllner 2020; Herrmann et al. 2008). Beugelsdijk et al. (2015) find that countries' scores on the Hofstede dimensions replicated from surveys have not changed much between birth cohorts, relative to the scores of other countries. Therefore, cross-country differences in culture are generally very stable over time, as theory would predict (see also Khesali et al. 2025). Among the six cultural dimensions identified by Hofstede, three are relevant for our study: individualism, power distance, and long-term orientation.

To test our fourth hypothesis, we control for Islam's influence on a society using the Muslim population share in 2010 as a proxy. Finally, in our analysis, we also consider a popular indicator of economically favorable cultural traits that was introduced by Tabellini (2008a). Generalized morality combines information on generalized trust and whether children ought to be taught tolerance and respect for others. As there is no strong theoretical link between generalized morality and constitutional compliance, a statistical relationship between them could be interpreted as suggesting that it is the general ability to cooperate and create economic prosperity that is promoted by generalized morality and, in turn, favors constitutional compliance.

Data on the Muslim population share is available for many more countries than the data from Hofstede et al. (2010) and Tabellini (2008a). To ensure comparability of our results across models, we limit our sample to countries for which at least one of the other cultural traits is observed. This results in a sample size of 115 observations for which information on the dependent variable is also available. However, some instrumental variable (IV) regression models are based on as few as 63 observations.

Instrumental variables regression

Our baseline models are estimated using OLS. Although cultural traits are generally very stable over time (Beugelsdijk et al. 2015; Khesali et al. 2025; Roland 2004; Williamson 2000), one might question whether a conditional correlation between constitutional compliance and present-day cultural traits allows for a causal interpretation. Here, we employ two additional identification strategies to address concerns about endogeneity. First, we employ two-stage least squares (2SLS) IV regression based on established external instruments for different cultural dimensions, which are introduced in the following paragraphs. Second, we augment the first approach by applying Lewbel's (2012) method for exploiting model heteroskedasticity to construct internal instruments using the available regressors in the model. These internal instruments can be used instead of external instruments, for example, when none are available. They can also be used to supplement external instruments, as will be the case here, in order to enhance estimation efficiency and to allow for formally testing the exclusion restriction when only one external instrument is available. Lewbel's method achieves identification by having regressors that are uncorrelated with the product of heteroskedastic errors, which is a feature of many models where error correlations are due to an unobserved common factor. This assumption is highly plausible in our application.

Gorodnichenko and Roland (2017, 2021) use two instruments to explain the global distribution of individualism: historical pathogen prevalence and blood distance, the Mahalanobis distance between the frequency of blood types in a country and their frequency in the United Kingdom. Fincher et al. (2008) argue that specific behavioral manifestations of collectivism inhibit the transmission of pathogens and, therefore, more often characterize cultures in areas with historically higher pathogen prevalence. Blood distance serves as a proxy for historical cultural distance, since genes and culture are simultaneously passed on from parents to their offspring and, thus, coevolve (see, e.g., Chiao and Blizinsky 2010). Importantly, Gorodnichenko and Roland (2017) explicitly address the possible concern that blood distance from the United Kingdom may capture geographical rather than cultural distance. They show that individualism remains robustly significant after controlling for the log of population-weighted geographical distance from the United Kingdom, while the geographical-distance variable itself is not statistically significant. Blood distance from the UK, one of the most individualistic countries, is therefore a suitable proxy for the level of intergenerationally transmitted individualism in a country.

To instrument power distance, we rely on an IV introduced by Gründler and Köllner (2020), which exploits regional patterns in the spatial distribution of culture. While spatial instruments are convenient, as they can easily be constructed based on any assumed spatial dependence patterns, their main drawback is their reliance on more demanding identifying assumptions (see, e.g., Betz et al. 2020). To construct the IV, Gründler and Köllner (2020) divide each continent into four disjoint (mutually exclusive) regions. The IV is then calculated for each country as the average level of power distance of all other countries located in the same region.

As IVs for long-term orientation, we rely on three geographic characteristics established by Galor and Özak (2016). The first two indicators capture the historical potential yield and duration of the growth cycle for the crop that maximizes potential yield on local land, measured in calories per hectare per year and averaged at the country level. In populations exposed to a higher crop yield, for a given growth cycle, the rewarding experience in agricultural investment should have triggered selection, adaptation, and learning processes which promoted the population's long-term orientation (Galor and Özak 2016). The third IV captures the years that elapsed since the onset of the Neolithic Revolution. A longer time since the onset of the Neolithic Revolution is associated with reduced long-term orientation, because the rise of institutionalized statehood after the transition to agriculture was associated with the taxation of crop yield and, thus, with reduced incentives to invest. All three IVs are ancestry adjusted to account for distortions caused by historical population movements. Galor and Özak (2016) provide evidence that crop yield is not related to any other standard metrics of culture, including individualism and power distance.

We instrument the Muslim population share with the distance between a country's capital and Mecca and an indicator for lasting historical conquest by Arab armies. Distance from Mecca is preferred here over the geographical distribution of historical trade routes, which Michalopoulos et al. (2018) claim has better predictive power, because the latter is potentially endogenous to various factors and, thus, less likely to meet the exclusion restriction as an IV. It has been argued that Arab conquest and the subsequent establishment of lasting Islamic empires were responsible for spreading the long-run legacy of control structures developed in the early Islamic world (Chaney 2012).

As instruments for generalized morality, we rely, as in the case of power distance, on IVs developed by Gründler and Köllner (2020). Specifically, we use two indicators of trust and tolerance in other countries in the same region. Tabellini (2008a) instead uses pronoun drop grammar rules as an IV for generalized morality. This approach has since drawn criticism (Murphy 2026) and pronoun drop is in any case an IV for individualism rather than for generalized morality (Kashima and Kashima 1998). Tabellini (2010) proposes historical executive constraints and literacy rates as alternative IVs, but he concedes that it is far from evident that the identifying assumptions hold.

Control variables

We also account for a set of factors that are both exogenous to a country's culture and potentially relevant to its level of constitutional compliance. We include four dummy variables for whether a country has been a colony for a long period of time and with significant influence of the colonial power in governance and whether it was, more specifically, under significant Spanish, French, or British colonial influence (Mayer and Zignago 2011).[5] Distinguishing different colonial powers is important, as they had fundamentally different colonization strategies with important consequences for political development in former colonies (e.g., Lange 2004). Moreover, we control for ethnic fractionalization as a general determinant of social accountability mechanisms that may force politicians to comply with the constitution (Alesina et al. 2003). Ethnic fractionalization affects the social fabric of society not via the presence of a particular cultural trait but through the diversity of the population. Finally, we add two binary variables proposed by Spolaore and Wacziarg (2013), identifying whether a country is on an island or whether it is landlocked. Both have been linked to cross-country differences in institutional quality (e.g., Carmignani 2015; Congdon Fors 2014).

Variable descriptions and summary statistics for all variables are provided in Table OA.1 in the Online Appendix and Table A.1 in the Appendix, respectively. Table A.2 shows the bivariate correlations between our culture variables. Table OA.2 lists the countries covered by our analysis and the levels of constitutional compliance and cultural characteristics for each country. Figure OA.1 visualizes the global distribution of our main dependent variable.

[5] Note that these dummies are not disjoint. Unlike indicators that only record the single most important colonial power, we also account for countries that had different major colonizers over time.

### 3.2 Estimation Results – Ordinary Least Squares

Table 1 presents our baseline estimation results without control variables, based on ordinary least squares (OLS) estimation and heteroskedasticity-robust standard errors. All coefficient estimates exhibit the theoretically expected sign. However, the correlation between generalized morality and constitutional compliance is the only one that is not statistically significant. As a robustness test, we estimate the same model in Table OA.3 using the robust regression estimator by Hamilton (1991), which corrects for influential observations, and the results are virtually identical. Our baseline results, thus, suggest that individualism, the absence of power distance, long-term orientation and a low Muslim population share are associated with higher levels of constitutional compliance.

Table 1: Effect of culture on compliance, no controls

| | (1) | (2) | (3) | (4) | (5) |
|---|---|---|---|---|---|
| | CC | CC | CC | CC | CC |
| Morality | 0.015 | | | | |
| | (0.109) | | | | |
| Individ. | | 0.221** | | | |
| | | (0.081) | | | |
| Power dist. | | | -0.254* | | |
| | | | (0.106) | | |
| Long-term | | | | 0.227* | |
| | | | | (0.111) | |
| Muslim | | | | | -0.518*** |
| | | | | | (0.070) |
| Controls | No | No | No | No | No |
| Estimator | OLS | OLS | OLS | OLS | OLS |
| Obs. | 79 | 101 | 67 | 89 | 115 |
| R-sq. | 0.00 | 0.05 | 0.08 | 0.05 | 0.27 |

Note: OLS regression coefficient estimates with robust standard errors in parentheses. Dependent variable: constitutional compliance. Constant omitted. + p<.1, * p<.05, ** p<.01, *** p<.001.

Table 2 shows the results of estimating the baseline model with control variables added. The key difference compared to Table 1 is that long-term orientation does not exhibit a statistically significant correlation with constitutional compliance after we include control variables. In terms of our control variables, we find little evidence that geography, colonial history, and fractionalization matter systematically for constitutional compliance. It may come as a surprise that specifically British colonial influence is sometimes associated with reduced constitutional compliance compared to other colonies. However, British colonies were exceptional in their use of indirect rule, under which political control was not centralized. Instead, tribalism was reinforced, parallel (modern and customary) legal systems were established, the market was replaced with customary possession of land, and no separation of powers existed on the local state level (Mamdani 1996, p. 22). Tribal customs were used to counter demands from the native population for equal treatment and equal civil status (p. 92).

In Tables OA.4 and OA.5, we further explore the influence of the Muslim population share, which is highly statistically significant in Column 5 of Tables 1 and 2. This ensures that what we observe is not the indirect effect of the presence of other major religions in countries where the Muslim population share is low. Table OA.4 considers the shares of Catholic and Protestant Christians and the Muslim population share one by one and then simultaneously. In Column 5, we also control for Orthodox Christians, Buddhists, and Hindus. Table OA.5 shows estimates based on (disjoint) dummy variables identifying whether an absolute majority of the population belongs to one of the same religious groups as in Table OA.4. The results in Tables OA.4 and OA.5 support that Islam has a robust and statistically significant negative association with constitutional compliance.

Table 2: Effect of culture on compliance, with controls

| | (1)<br>CC | (2)<br>CC | (3)<br>CC | (4)<br>CC | (5)<br>CC |
|---|---|---|---|---|---|
| Morality | 0.031 | | | | |
| | (0.130) | | | | |
| Individ. | | 0.333** | | | |
| | | (0.111) | | | |
| Power dist. | | | -0.349** | | |
| | | | (0.120) | | |
| Long-term | | | | 0.200 | |
| | | | | (0.157) | |
| Muslim | | | | | -0.539*** |
| | | | | | (0.083) |
| Island | 0.217 | 0.184 | 0.037 | 0.199 | -0.024 |
| | (0.242) | (0.250) | (0.274) | (0.254) | (0.200) |
| Landlocked | -0.273 | 0.274 | 0.186 | -0.327 | -0.379 |
| | (0.320) | (0.286) | (0.335) | (0.290) | (0.262) |
| Colony: ESP | -0.074 | 0.120 | -0.267 | 0.146 | -0.401+ |
| | (0.305) | (0.224) | (0.216) | (0.334) | (0.224) |
| Colony: FRA | -0.644 | -0.884** | -0.819 | -0.267 | -0.159 |
| | (0.474) | (0.329) | (0.633) | (0.414) | (0.321) |
| Colony: GBR | -0.809* | -0.691** | -0.942** | -0.504 | -0.512* |
| | (0.309) | (0.254) | (0.338) | (0.308) | (0.199) |
| Colonized | 0.277 | 0.682* | 0.823** | 0.264 | 0.307 |
| | (0.328) | (0.281) | (0.282) | (0.285) | (0.226) |
| Fractional. | -0.073 | -0.189 | -0.161 | 0.142 | 0.086 |
| | (0.560) | (0.466) | (0.591) | (0.489) | (0.369) |
| Estimator | OLS | OLS | OLS | OLS | OLS |
| Obs. | 79 | 101 | 67 | 89 | 115 |
| R-sq. | 0.13 | 0.21 | 0.31 | 0.11 | 0.32 |

Note: OLS regression coefficient estimates with robust standard errors in parentheses. Dependent variable: constitutional compliance. Constant omitted. + p<.1, * p<.05, ** p<.01, *** p<.001.

Protestantism is the only other religion with a significant but not fully robust correlation with constitutional compliance. A stronger influence of Protestantism on society is associated with a higher compliance level. This is not surprising, as already Durkheim (1897) pointed out that Protestant doctrine encourages independent thought and religious individualism (see also Becker and Woessmann 2018). Therefore, Protestantism may function here merely as an alternative individualism indicator, which so far exhibits a positive association with constitutional compliance in line with H1.

### 3.3 Estimation Results – Instrumental Variables Regression

Thus far, we have discussed ordinary least squares estimates. Although a causal interpretation of these conditional correlations is not implausible, as discussed in Section 2, we have avoided causal language. We now compare our baseline results to those derived from instrumental variables estimation, using the IVs introduced and rationalized in Section 3.1. Table 3 shows the results of 2SLS IV regression. As before, all coefficients exhibit the theoretically expected signs in line with Hypotheses 1 to 4. However, all coefficient estimates in Table 3 are statistically significant. None of the reported Hansen J statistics indicate a violation of the exclusion restriction. A test of the exclusion restriction is not possible for power distance, as the model is exactly identified, i.e., the number of IVs is equal to the number of endogenous regressors. The model for long-term orientation in Column 4 is the only one showing signs of weak instruments. Its first-stage F-statistic is below the Stock-Yogo critical value for 20% maximal IV size (Stock and Yogo 2005). It is also below the less refined and indiscriminate rule of thumb proposed by Staiger and Stock (1997) that the F-statistic should exceed 10. In Table 3 and all the following regression tables, we report where possible weak-IV robust confidence sets as proposed by Andrews (2018) and calculated following Sun (2018). These allow for reliable inference, even if the F-statistic is below the corresponding Stock-Yogo critical value. If that is not the case, the weak-IV robust confidence set may still be of interest, as the Stock-Yogo values are calculated under the assumption of homoskedasticity. The weak-IV robust confidence set in Column 4 does not enclose 0, suggesting that long-term orientation does indeed have a statistically significant positive effect on compliance.

Table 3: Effect of culture on compliance, 2SLS

| | (1) CC | (2) CC | (3) CC | (4) CC | (5) CC |
|---|---|---|---|---|---|
| Morality | 0.369* | | | | |
| | (0.161) | | | | |
| Individ. | | 0.684*** | | | |
| | | (0.184) | | | |
| Power dist. | | | -0.737* | | |
| | | | (0.294) | | |
| Long-term | | | | 0.795* | |
| | | | | (0.333) | |
| Muslim | | | | | -0.700*** |
| | | | | | (0.089) |
| Controls | Yes | Yes | Yes | Yes | Yes |
| Estimator | 2SLS | 2SLS | 2SLS | 2SLS | 2SLS |
| Obs. | 68 | 98 | 63 | 86 | 115 |
| R-sq. | 0.04 | 0.11 | 0.11 | -0.08 | 0.30 |
| F-stat | 17.08 | 27.73 | 11.79 | 6.61 | 102.17 |
| Stock-Yogo | 8.75 | 8.75 | 6.66 | 9.54 | 8.75 |
| Robust CI | [.09],[.81] | [.38],[1.10] | [-1.64],[-.30] | [.24],[1.81] | [-.88],[-.53] |
| J-stat | 1.18 | 0.98 | . | 4.28 | 0.35 |

Note: 2SLS regression coefficient estimates with robust standard errors in parentheses. Dependent variable: constitutional compliance. “F-stat” refers to the Kleibergen-Paap rk Wald F statistic. Stock and Yogo (2005)-critical values for weak identification are reported for 20% maximal IV size. Instruments are considered weak if the F-test falls below the “Stock-Yogo” critical value. “Robust CI” refers to an identification-robust confidence set of the independent (culture) variable of interest that is robust to weak instruments (see Andrews 2018; Sun 2018). The weak-IV robust estimate is significant at the 5%-level if the confidence set does not enclose zero. “J-stat” refers to the Sargan-Hansen test of overidentifying restrictions. A * on J-stat indicates that the null hypothesis of instrument validity can be rejected at the 5%-level. Column 3 does not report a J-statistic, as the model is exactly identified. Constant omitted. + p<.1, * p<.05, ** p<.01, *** p<.001.

In Table 4, we reproduce the results from Table 3 with the same external instruments and adding internal instruments constructed following the Lewbel method. These heteroskedasticity-based internal IVs can increase the estimation efficiency and allow us to test the exclusion restriction in all models.

Table 4: Effect of culture on compliance, Lewbel estimator

| | (1) CC | (2) CC | (3) CC | (4) CC | (5) CC |
|---|---|---|---|---|---|
| Morality | 0.059 | | | | |
| | (0.133) | | | | |
| Individ. | | 0.594*** | | | |
| | | (0.147) | | | |
| Power dist. | | | -0.360+ | | |
| | | | (0.199) | | |
| Long-term | | | | 0.240 | |
| | | | | (0.257) | |
| Muslim | | | | | -0.614*** |
| | | | | | (0.090) |
| Controls | Yes | Yes | Yes | Yes | Yes |
| Estimator | Lewbel | Lewbel | Lewbel | Lewbel | Lewbel |
| Obs. | 68 | 98 | 63 | 86 | 115 |
| R-sq. | 0.13 | 0.15 | 0.28 | 0.11 | 0.32 |
| F-stat | 12.36 | 7.84 | 4.53 | 5.45 | 86.90 |
| Stock-Yogo | 6.65 | 6.65 | 6.69 | 6.61 | 6.65 |
| Robust CI | . | [.48],[1.12] | . | [-1.48],[1.42] | [-.82],[-.54] |
| J-stat | 14.55 | 7.52 | 10.50 | 18.39* | 10.35 |

Note: 2SLS regression coefficient estimates with robust standard errors in parentheses. Dependent variable: constitutional compliance. External instruments are supplemented with heteroskedasticity-based instrumental variables constructed using Lewbel's (2012) method (see Baum and Schaffer 2026). "F-stat" refers to the Kleibergen-Paap rk Wald F statistic. Stock and Yogo (2005)-critical values for weak identification are reported for 20% maximal IV relative bias. Instruments are considered weak if the F-test falls below the "Stock-Yogo" critical value. "Robust CI" refers to an identification-robust confidence set of the independent (culture) variable of interest that is robust to weak instruments (see Andrews 2018; Sun 2018). The weak-IV robust estimate is significant at the 5%-level if the confidence set does not enclose zero. "J-stat" refers to the Sargan-Hansen test of overidentifying restrictions. A * on J-stat indicates that the null hypothesis of instrument validity can be rejected at the 5%-level. In Columns 1 and 3, the weak-IV robust procedure does not yield a single bounded interval. Constant omitted. + p<.1, * p<.05, ** p<.01, *** p<.001.

All models reported in Table 4 except that using long-term orientation pass the exclusion restriction test. Moreover, the estimated effect of power distance and long-term orientation suffer from weak instruments, and for generalized morality and power distance, the weak-IV robust procedure does not yield a single bounded interval. Considering the results in Table 4 and adopting a conservative interpretation, our findings suggest that only two cultural traits have a robust effect. Increasing individualism by one standard deviation or decreasing the Muslim population share by one standard deviation, respectively, increases constitutional compliance by more than half a standard deviation.

### 3.4 Estimation Results – Extensions and Robustness

In the remainder of the empirical analysis, we want to answer five questions regarding the effects of individualism and Islam on constitutional compliance: (1) Does their effect operate (exclusively) via their effect on democracy or do they have an independent effect? (2) Do the two cultural traits affect constitutional compliance in particular legal areas or across the board? (3) Do individualism and Islam have separate effects, although they are (negatively) correlated with each other? (4) Are other cultural dimensions, which are frequently linked to individualism and Islam, correlated with constitutional compliance? And (5) do the different cultural dimensions exert their influence on constitutional compliance through strengthening civil society or independent veto players? Or do they have a direct effect (i.e., independent from these two channels)?

In Tables A.3 and OA.6, we re-estimate the models in Tables 3 and 4 while controlling for democracy, based on the minimalist indicator by Bjørnskov and Rode (2020). Given that individualism (Gorodnichenko and Roland 2021) and Islam (Gouda and Hanafy 2022; Potrafke 2012, 2013) have already been linked to democracy, and that democracy, in turn, is conducive to constitutional compliance (Gutmann et al. 2024b), this analysis can clarify if there is an independent effect of culture on constitutional compliance or if the effect works only indirectly via promoting democracy. Applying a minimalist definition of electoral democracy here is critical to prevent the democracy indicator from measuring elements of constitutional compliance itself, for example, in the form of executive constraints or the strength of *de facto* political rights. The (direct) effects of individualism and Islam shrink by about 40% to 50%, but they remain statistically significant.[6] The democracy indicator itself also exhibits a statistically significant correlation. Democracies have a level of constitutional compliance that is more than one standard deviation higher than in nondemocracies.

In Tables A.4 and A.5, we reproduce the estimates from Columns 2 and 5 in Table 4. However, we swap the dependent variable of overall constitutional compliance for the four constitutional compliance indicators for the specific legal domains (1) property rights and the rule of law, (2) political rights, (3) civil liberties, and (4) basic human rights. We find significant effects of both individualism and Islam on all four dimensions. The effect sizes are also very similar, except the effect of Islam on compliance with constitutional basic human rights, which is about half the effect size compared to the other dependent variables. Table OA.7 presents the same analysis, but with the effects of individualism and Islam measured in the same model. Both exert a statistically significant effect on overall constitutional compliance (Column 1). The effects of individualism and Islam on compliance in the four legal areas are statistically significant, except for the effects of individualism on compliance with political and civil rights in the constitution. Weak-IV robust confidence sets suggest statistical significance of all coefficient estimates. We interpret these results as clearly suggesting that both individualism and Islam have their own effect on governments' compliance with the national constitution. The effect of individualism appears most pronounced in the domain of property rights and the rule of law. The effect of Islam, in contrast, is a bit more uniformly distributed.

[6] The model for individualism based on the Lewbel estimator appears to suffer from weak IVs, but the weak-IV robust confidence set confirms that the effect of individualism is significant at the 5%-level.

Next, we take a closer look at two additional well-established cultural dimensions: “kinship tightness” and “tightness/looseness”. They are related to but distinct from individualism and Islam, regarding their association with constitutional compliance. For this extension, we estimate a set of models as in Table 2. The results are presented in Table OA.8. First, we rely on an indicator by Enke (2019) for the historical prevalence of kinship tightness in preindustrial societies. It captures the extent to which people were interconnected in tightly structured, extended family systems. The historical prevalence of tight kinship favors communal values at the expense of universalistic values that would allow cooperation with strangers outside the kinship structure (Enke 2019). Hence, countries in which the population’s ancestors relied on tight kinship should be less individualistic to this day (see also Schulz et al. 2019). Henrich et al. (2010) stress the importance of kinship tightness for the broader concept of WEIRD (western, educated, industrialized, rich, and democratic) populations. They argue that the moral reasoning based on abstract ethical principles regarding justice and individual rights and inherent in most Western constitutions can only be commonly found in WEIRD populations. We find that a one standard deviation higher historical kinship tightness is associated with a one-third of a standard deviation decline in constitutional compliance. The estimated coefficient and standard error are almost identical with those reported for individualism in Table 2. We also measure the conditional correlation between constitutional compliance and tightness, as in the cultural tightness-looseness dimension. Superficially, one might expect that tight societies, which are characterized by having many strong norms and a low tolerance of deviant behavior, also expect their government to comply with the constitution. However, these societies tend to exhibit rule by law rather than rule of law. Although these societies have many laws and strict law enforcement (Gelfand et al. 2011), it is not clear that political leaders are held to the same standard as citizens. Using tightness data by Gelfand et al. (2011) and Eriksson et al. (2021), we find no significant association with constitutional compliance.

Finally, in Section 2, we have argued that two of the main actors that can link cultural traits to constitutional compliance are civil society and political veto players. In an attempt to corroborate that these actors play a role in the causal mechanisms linking culture to compliance, we conduct a causal mediation analysis (Daniel et al. 2015), relying on a generalized method of moments estimator. The strength of civil society is measured as the core civil society index from V-Dem, whereas the strength of veto players is proxied by the level of independence of the judiciary. Table OA.9 presents the decomposed effect estimates.

We find that long-term orientation (positive) and Islam (negative) exert a significant effect on constitutional compliance that is not mediated through civil society or independent veto players. The only cultural trait that significantly affects constitutional compliance via veto player strength (as proxied by judicial independence) is individualism. In contrast, mediation via civil society strength occurs for individualism, power-distance, and Islam. Given that only the effects of individualism and Islam have turned out to be robust in our previous analysis, the results of the causal mediation analysis can be summarized as follows: While the positive effect of individualism on constitutional compliance is mediated by both civil society and veto players, the negative effect of Islam is primarily mediated via civil society and otherwise occurs independent of the potential mediators studied here.

## 4. Conclusion

Constitutions can be interpreted as sets of promises. If constitutional provisions are reliably enforced, they are conducive to economic and political stability and can help citizens form accurate expectations. This is a crucial precondition for long-term investments, which, in turn, promote growth and development. Constitutional compliance is, therefore, of high value to a society.

The results of our empirical study show that culture is relevant for constitutional compliance. More individualistic societies have governments that comply more with their national constitution. More Islamic societies, in the sense that they have a higher Muslim population share, experience significantly less constitutional compliance. These effects are largely independent of each other and not exclusively mediated via the effect of culture on democracy. Other cultural dimensions—specifically power distance, long-term orientation, and generalized morality—are not robustly related to constitutional compliance levels. Overall, our results can be interpreted as evidence that constitutional compliance constitutes a transmission channel between cultural factors and economic development.

Beyond a better understanding of the long-run causes of economic growth, our results can be interpreted as providing a warning that the effectiveness of formal constitutional rules (and by extension legislation) depends at least to some extent on the cultural context in which they are embedded. The literature on legal transplants has emphasized that formal laws need to match the setting in which they are implemented. We, in contrast, provide evidence that some cultural traits are conducive or detrimental to constitutional compliance, largely independent of the constitutional rules in question.

Future research might inquire further into how a mismatch between the choice of constitutional rules and national culture affects constitutional compliance, especially with those mismatched rules. If negative effects are observable, it is important to understand why mismatched rules are adopted in the first place. One possible explanation is that in the past, some constitution-makers simply copied successful constitutions, without considering that their effects are conditional on the cultural and social fabric of a society. Political and economic institutions that are effective in a high-trust environment, for example, might quickly fail when adopted by a low-trust society (Djankov et al. 2003). A second possible explanation for the adoption of poorly adapted constitutional rules could be the technical assistance provided by international organizations, such as UNDP. Finally, micro-level survey evidence could complement our macro-level analysis by shedding light on how individual values relate to attitudes toward constitutional compliance.

# Appendix

Table A.1: Descriptive statistics

| | N | Mean | SD | Min | Max |
|---|---|---|---|---|---|
| CC | 115 | 0 | 1 | -2.16 | 1.39 |
| Prop | 115 | 0 | 1 | -1.76 | 1.45 |
| Polit | 114 | 0 | 1 | -2.16 | 1.26 |
| Civil | 115 | 0 | 1 | -2.38 | 0.99 |
| Basic | 115 | 0 | 1 | -2.54 | 1.12 |
| Morality | 79 | 0 | 1 | -1.89 | 2.93 |
| Individ. | 101 | 0 | 1 | -1.54 | 2.28 |
| Power dist. | 67 | 0 | 1 | -2.27 | 2.07 |
| Long-term | 89 | 0 | 1 | -1.77 | 2.25 |
| Muslim | 115 | 0 | 1 | -0.66 | 2.24 |
| Kinship tightness | 98 | 0 | 1 | -1.61 | 1.59 |
| Tightness (32) | 30 | 0 | 1 | -1.75 | 2.00 |
| Tightness (57) | 52 | 0 | 1 | -1.49 | 2.88 |
| Island | 115 | 0.14 | 0.35 | 0 | 1 |
| Landlocked | 115 | 0.20 | 0.40 | 0 | 1 |
| Colony: ESP | 115 | 0.13 | 0.34 | 0 | 1 |
| Colony: FRA | 115 | 0.11 | 0.32 | 0 | 1 |
| Colony: GBR | 115 | 0.25 | 0.44 | 0 | 1 |
| Colonized | 115 | 0.77 | 0.43 | 0 | 1 |
| Fractional. | 115 | 0.43 | 0.24 | 0.00 | 0.93 |
| Democracy | 115 | 0.69 | 0.47 | 0 | 1 |
| Veto players | 115 | 0.80 | 1.27 | -2.07 | 3.19 |
| Civil society | 115 | 0.75 | 0.24 | 0.04 | 0.98 |
| Catholic share | 115 | 0.28 | 0.31 | 0.00 | 0.96 |
| Protestant share | 115 | 0.13 | 0.19 | 0.00 | 0.87 |
| Muslim share | 115 | 0.23 | 0.34 | 0.00 | 0.99 |
| Orthodox share | 115 | 0.10 | 0.24 | 0.00 | 0.95 |
| Buddhist share | 115 | 0.04 | 0.16 | 0.00 | 0.87 |
| Hindu share | 115 | 0.03 | 0.11 | 0.00 | 0.81 |
| Catholic majority | 115 | 0.27 | 0.45 | 0 | 1 |
| Protestant majority | 115 | 0.05 | 0.22 | 0 | 1 |
| Muslim majority | 115 | 0.22 | 0.41 | 0 | 1 |
| Orthodox majority | 115 | 0.10 | 0.30 | 0 | 1 |
| Buddhist majority | 115 | 0.03 | 0.18 | 0 | 1 |
| Hindu majority | 115 | 0.02 | 0.13 | 0 | 1 |
| Trust, Spatial | 68 | 27.34 | 12.20 | 6.50 | 50.50 |
| Tolerance, Spatial | 68 | 67.29 | 5.74 | 54.40 | 77.26 |
| Blood dist. UK | 98 | 1.57 | 0.83 | 0.00 | 3.59 |
| Pathogen prev. | 101 | 0.02 | 0.66 | -1.31 | 1.16 |
| Power dist., Spatial | 63 | 59.78 | 15.32 | 30.36 | 84.25 |
| Crop yield | 88 | 0.01 | 1.00 | -3.44 | 2.30 |
| Crop growth cycle | 88 | -0.01 | 1.01 | -6.38 | 2.54 |
| Neolithic trans. | 86 | -0.00 | 0.99 | -2.38 | 2.35 |
| Arab conquest | 115 | 0.13 | 0.33 | 0.00 | 1.00 |
| Dist. to Mecca | 115 | 1.55 | 0.65 | -0.24 | 2.75 |

Table A.2: Correlations between culture indicators

| | Morality | Individ. | Power dist. | Long-term | Muslim |
|---|---|---|---|---|---|
| Morality | 1.00 | | | | |
| Individ. | 0.49 | 1.00 | | | |
| Power dist. | -0.60 | -0.63 | 1.00 | | |
| Long-term | 0.01 | 0.15 | 0.03 | 1.00 | |
| Muslim | -0.04 | -0.22 | 0.20 | -0.22 | 1.00 |

Note: Pearson correlation coefficients.

Table A.3: Effect independent of democracy, 2SLS

| | (1) | (2) | (3) | (4) | (5) |
|---|---|---|---|---|---|
| | CC | CC | CC | CC | CC |
| Morality | 0.048 | | | | |
| | (0.104) | | | | |
| Individ. | | 0.410* | | | |
| | | (0.163) | | | |
| Power dist. | | | -0.447 | | |
| | | | (0.334) | | |
| Long-term | | | | 0.298 | |
| | | | | (0.272) | |
| Muslim | | | | | -0.434*** |
| | | | | | (0.114) |
| Democracy | 1.594*** | 1.206*** | 1.319* | 1.421*** | 1.096*** |
| | (0.268) | (0.238) | (0.547) | (0.242) | (0.230) |
| Controls | Yes | Yes | Yes | Yes | Yes |
| Estimator | 2SLS | 2SLS | 2SLS | 2SLS | 2SLS |
| Obs. | 68 | 98 | 63 | 86 | 115 |
| R-sq. | 0.50 | 0.43 | 0.43 | 0.42 | 0.53 |
| F-stat | 14.74 | 21.48 | 7.37 | 6.39 | 61.33 |
| Stock-Yogo | 8.75 | 8.75 | 6.66 | 9.54 | 8.75 |
| Robust CI | [-.15],[.33] | [.14],[.77] | [-1.80],[.05] | . | [-.67],[-.22] |
| J-stat | 3.69 | 2.12 | . | 4.78 | 0.01 |

Note: 2SLS regression coefficient estimates with robust standard errors in parentheses. Dependent variable: constitutional compliance. "F-stat" refers to the Kleibergen-Paap rk Wald F statistic. Stock and Yogo (2005)-critical values for weak identification are reported for 20% maximal IV size. Instruments are considered weak if the F-test falls below the "Stock-Yogo" critical value. "Robust CI" refers to an identification-robust confidence set of the independent (culture) variable of interest that is robust to weak instruments (see Andrews 2018; Sun 2018). The weak-IV robust estimate is significant at the 5%-level if the confidence set does not enclose zero. "J-stat" refers to the Sargan-Hansen test of overidentifying restrictions. A * on J-stat indicates that the null hypothesis of instrument validity can be rejected at the 5%-level. Column 3 does not report a J-statistic, as the model is exactly identified. In Column 4, the weak-IV robust procedure does not yield a single bounded interval. Constant omitted. + p<.1, * p<.05, ** p<.01, *** p<.001.

Table A.4: Effect of individualism on compliance dimensions

| | (1) Prop | (2) Polit | (3) Civil | (4) Basic |
|---|---|---|---|---|
| Individ. | 0.604*** | 0.474*** | 0.525*** | 0.454** |
| | (0.148) | (0.127) | (0.154) | (0.140) |
| Controls | Yes | Yes | Yes | Yes |
| Estimator | Lewbel | Lewbel | Lewbel | Lewbel |
| Obs. | 98 | 97 | 98 | 98 |
| R-sq. | 0.24 | 0.20 | 0.15 | 0.04 |
| F-stat | 7.84 | 7.70 | 7.84 | 7.84 |
| Stock-Yogo | 6.65 | 6.65 | 6.65 | 6.65 |
| Robust CI | [.44],[1.16] | [.38],[.87] | [.35],[.98] | [.30],[.92] |

Note: 2SLS regression coefficient estimates with robust standard errors in parentheses. Dependent variables: constitutional compliance dimensions. External instruments are supplemented with heteroskedasticity-based instrumental variables constructed using Lewbel's (2012) method (see Baum and Schaffer 2026). "F-stat" refers to the Kleibergen-Paap rk Wald F statistic. Stock and Yogo (2005)-critical values for weak identification are reported for 20% maximal IV relative bias. Instruments are considered weak if the F-test falls below the "Stock-Yogo" critical value. "Robust CI" refers to an identification-robust confidence set of the independent (culture) variable of interest that is robust to weak instruments (see Andrews 2018; Sun 2018). The weak-IV robust estimate is significant at the 5%-level if the confidence set does not enclose zero. Constant omitted. + p<.1, * p<.05, ** p<.01, *** p<.001.

Table A.5: Effect of Muslim share on compliance dimensions

| | (1) | (2) | (3) | (4) |
|---|---|---|---|---|
| | Prop | Polit | Civil | Basic |
| Muslim | -0.553*** | -0.564*** | -0.683*** | -0.333** |
| | (0.081) | (0.093) | (0.100) | (0.114) |
| Controls | Yes | Yes | Yes | Yes |
| Estimator | Lewbel | Lewbel | Lewbel | Lewbel |
| Obs. | 115 | 114 | 115 | 115 |
| R-sq. | 0.31 | 0.32 | 0.35 | 0.15 |
| F-stat | 86.90 | 84.83 | 86.90 | 86.90 |
| Stock-Yogo | 6.65 | 6.65 | 6.65 | 6.65 |
| Robust CI | [-.67],[-.46] | [-.79],[-.47] | [-.96],[-.60] | [-.59],[-.14] |

Note: 2SLS regression coefficient estimates with robust standard errors in parentheses. Dependent variables: constitutional compliance dimensions. External instruments are supplemented with heteroskedasticity-based instrumental variables constructed using Lewbel's (2012) method (see Baum and Schaffer 2026). "F-stat" refers to the Kleibergen-Paap rk Wald F statistic. Stock and Yogo (2005)-critical values for weak identification are reported for 20% maximal IV relative bias. Instruments are considered weak if the F-test falls below the "Stock-Yogo" critical value. "Robust CI" refers to an identification-robust confidence set of the independent (culture) variable of interest that is robust to weak instruments (see Andrews 2018; Sun 2018). The weak-IV robust estimate is significant at the 5%-level if the confidence set does not enclose zero. Constant omitted. + p<.1, * p<.05, ** p<.01, *** p<.001.

# Online Appendix for "Culture and constitutional compliance"

Figure OA.1: Global distribution of constitutional compliance

Note: Displayed are constitutional compliance values for the 115 countries in our sample. Darker blue colors indicate higher levels of constitutional compliance.

Table OA.1: Variable description and data sources

| Variable name | Description | Data Source |
|---|---|---|
| CC | Constitutional compliance: Extent to which *de jure* constitutional rules are complied with *de facto* by the government. Based on 14 types of constitutional rules representing four legal dimensions, mean value 2010-2019. | Gutmann et al. (2024b), v2.0 |
| Prop | Constitutional compliance, subdimension: property rights and the rule of law (e.g., private property rights), mean value 2010-2019. | Gutmann et al. (2024b), v2.0 |
| Polit | Constitutional compliance, subdimension: political rights (e.g., freedom of assembly), mean value 2010-2019. | Gutmann et al. (2024b), v2.0 |
| Civil | Constitutional compliance, subdimension: civil rights (e.g., freedom of speech), mean value 2010-2019. | Gutmann et al. (2024b), v2.0 |
| Basic | Constitutional compliance, subdimension: basic human rights (e.g., prohibition of torture), mean value 2010-2019. | Gutmann et al. (2024b), v2.0 |
| Morality | Generalized morality: First principal component of generalized trust and tolerance towards others (standardized). | Tabellini (2008a) |
| Individ. | Individualism: The extent to which people feel independent, as opposed to being interdependent as members of larger wholes (standardized). | Hofstede et al. (2010) |
| Power dist. | Power distance: The extent to which the less powerful members of groups or organizations accept and expect that power is distributed unequally (standardized). | Hofstede et al. (2010) |
| Long-term | Long-term orientation: Extent to which the world is thought of as being flux and preparing for the future is always needed. In a short-time-oriented culture, the past | Hofstede et al. (2010) |
| | provides a moral compass and adhering to it is morally good (standardized). | |
| Muslim | Muslim population share, 2010 (standardized). | Maoz and Henderson (2013) |
| Kinship tightness | Extent to which social organization is based on strong kinship ties and intensive family (standardized). | Enke (2019) |
| Tightness (32) | Extent to which social norms are pervasive and strictly enforced, as opposed to more permissive and weakly enforced norms (standardized). | Gelfand (2011) |
| Tightness (57) | Extent to which social norms are pervasive and strictly enforced, as opposed to more permissive and weakly enforced norms (standardized). | Eriksson (2021) |
| Island | Country is an island (1=yes, 0=no). | Spolaore and Wacziarg (2013) |
| Landlocked | Country is landlocked (1=yes, 0=no). | Spolaore and Wacziarg (2013) |
| Colony: ESP | Spain was a colonizer over a long period and with substantial participation in governance (1=yes, 0=no). | Mayer and Zignago (2011) |
| Colony: FRA | France was a colonizer over a long period and with substantial participation in governance (1=yes, 0=no). | Mayer and Zignago (2011) |
| Colony: GBR | Great Britain was a colonizer over a long period and with substantial participation in governance (1=yes, 0=no). | Mayer and Zignago (2011) |
| Colonized | Colonized over a long period and with substantial participation in governance by at least one colonial power (1=yes, 0=no). | Mayer and Zignago (2011) |
| Fractional. | Probability that two randomly selected individuals from the population belong to different ethnic groups. | Alesina et al. (2003) |
| Democracy | Electoral democracy, 2010 (1=yes, 0=no). | Bjørnskov and Rode (2020) |
| Veto players | Mean value of high court (v2juhcind) and | V-Dem, v16 |
| | lower court independence (v2juncind) | |
| Civil society | Core civil society index (v2xcs_ccsi) | V-Dem, v16 |
| Catholic share | Catholic population share, 2010. | Maoz and Henderson (2013) |
| Protestant share | Protestant population share, 2010. | Maoz and Henderson (2013) |
| Muslim share | Muslim population share, 2010. | Maoz and Henderson (2013) |
| Orthodox share | Orthodox population share, 2010. | Maoz and Henderson (2013) |
| Buddhist share | Buddhist population share, 2010. | Maoz and Henderson (2013) |
| Hindu share | Hindu population share, 2010. | Maoz and Henderson (2013) |
| Catholic majority | Over 50% Catholic, 2010 (1=yes, 0=no). | Maoz and Henderson (2013) |
| Protestant majority | Over 50% Protestant, 2010 (1=yes, 0=no). | Maoz and Henderson (2013) |
| Muslim majority | Over 50% Muslim, 2010 (1=yes, 0=no). | Maoz and Henderson (2013) |
| Orthodox majority | Over 50% Orthodox, 2010 (1=yes, 0=no). | Maoz and Henderson (2013) |
| Buddhist majority | Over 50% Buddhist, 2010 (1=yes, 0=no). | Maoz and Henderson (2013) |
| Hindu majority | Over 50% Hindu, 2010 (1=yes, 0=no). | Maoz and Henderson (2013) |
| Trust, Spatial | Mean generalized trust level among the countries in the same geographic region. | Gründler and Köllner (2020) |
| Tolerance, Spatial | Mean tolerance level among the countries in the same geographic region. | Gründler and Köllner (2020) |
| Blood dist. UK | Mahalanobis distance of the frequency of blood types A and B relative to the frequency of these blood types in the UK. | Gorodnichenko and Roland (2017) |
| Pathogen prev. | 9-item index of historical disease prevalence. | Murray and Schaller (2010) |
| Power dist., Spatial | Mean power distance level among the countries in the same geographic region. | Gründler and Köllner (2020) |
| Crop yield | Potential crop yield, adjusted for ancestral population movements. | Galor and Özak (2016) |
| Crop growth cycle | Potential crop growth cycle, adjusted for ancestral population movements. | Galor and Özak (2016) |
| Neolithic trans. | Number of years since the onset of the Neolithic Revolution, adjusted for ancestral population movements. | Ashraf and Galor (2011) and Galor and Özak (2016) |
| Arab conquest | Share of a country's landmass that was ruled by Muslim dynasties in 1100 CE. Set to zero for Israel or if less than half of the country's landmass was under Muslim rule | Chaney (2012) |
| | in 1900 CE. | |
| Dist. to Mecca | Natural log. of the distance between the country's capital and Mecca in 1,000 km. | Gouda and Gutmann (2021) |

Note: Standardized variables are subject to linear transformation such that their mean value is zero and their standard deviation is one, which simplifies the comparison of effect sizes.

Table OA.2: List of countries and main variables

| Country | CC | Morality | Individ. | Power dist. | Long-term | Muslim |
|---|---|---|---|---|---|---|
| Albania | 0.548 | 0.586 | -0.911 | . | 0.613 | 1.186 |
| Algeria | -1.039 | -1.661 | . | . | -0.852 | 2.242 |
| Angola | -1.451 | . | -1.001 | . | . | -0.631 |
| Argentina | 0.617 | -0.453 | 0.256 | -0.499 | -1.103 | -0.617 |
| Armenia | 0.345 | -1.223 | . | . | 0.613 | -0.660 |
| Australia | -0.265 | 1.402 | 2.232 | -1.013 | -1.062 | -0.596 |
| Austria | 0.042 | 0.293 | 0.660 | -2.273 | 0.571 | -0.522 |
| Azerbaijan | -1.766 | -0.864 | . | . | 0.613 | 2.125 |
| Bangladesh | -1.643 | -0.113 | -0.911 | 0.949 | 0.027 | 1.970 |
| Belarus | -1.394 | 0.063 | . | . | 1.450 | -0.655 |
| Belgium | 0.724 | 0.230 | 1.558 | 0.248 | 1.492 | -0.515 |
| Benin | 0.597 | . | 1.693 | . | . | 0.115 |
| Bhutan | 0.134 | . | 0.525 | . | . | -0.632 |
| Bosnia and Herzegovina | 0.804 | . | . | . | 0.990 | 0.659 |
| Brazil | 0.713 | -1.484 | -0.103 | 0.435 | -0.099 | -0.661 |
| Bulgaria | 0.961 | -0.873 | -0.462 | 0.482 | 0.948 | -0.283 |
| Burkina Faso | 0.708 | . | -1.136 | . | -0.810 | 1.037 |
| Canada | 0.536 | 1.422 | 1.783 | -0.966 | -0.434 | -0.604 |
| Central African Republic | -1.639 | . | 1.468 | . | . | -0.221 |
| Chile | 1.009 | 0.071 | -0.777 | 0.155 | -0.643 | -0.660 |
| China | -1.748 | 1.008 | -0.911 | 0.949 | 1.701 | -0.588 |
| Colombia | 0.357 | -0.862 | -1.226 | 0.341 | -1.396 | -0.660 |
| Costa Rica | 1.271 | . | -1.136 | -1.153 | . | -0.661 |
| Croatia | 1.244 | -0.377 | -0.328 | 0.622 | 0.487 | -0.618 |
| Czech Republic | 1.390 | -0.203 | 0.795 | -0.125 | 0.990 | -0.661 |
| Denmark | 0.518 | 2.105 | 1.513 | -1.947 | -0.475 | -0.544 |
| Dominican Republic | 0.156 | -0.039 | -0.462 | . | -1.396 | -0.661 |
| Ecuador | -0.168 | . | -1.450 | 0.855 | . | -0.661 |
| Egypt | -1.999 | 0.501 | -0.103 | . | -1.648 | 1.874 |
| El Salvador | 0.166 | -1.197 | -0.956 | 0.295 | -1.103 | -0.660 |
| Estonia | 1.289 | -0.181 | 0.885 | -0.919 | 1.492 | -0.657 |
| Ethiopia | -1.855 | . | -0.597 | . | . | 0.349 |
| Fiji | -0.641 | . | -1.181 | . | . | -0.485 |
| Finland | 1.081 | 2.443 | 1.019 | -1.246 | -0.350 | -0.648 |
| France | -0.229 | 0.144 | 1.379 | 0.388 | 0.697 | -0.430 |
| Georgia | 0.254 | -1.247 | . | . | -0.350 | -0.371 |
| Germany | 1.089 | 0.868 | 1.199 | -1.153 | 1.534 | -0.518 |
| Ghana | 1.017 | . | -0.911 | . | -1.773 | -0.145 |
| Greece | 1.170 | -1.168 | -0.238 | 0.015 | -0.057 | -0.591 |
| Guatemala | 0.095 | . | -1.540 | 1.649 | . | -0.661 |
| Honduras | -0.023 | . | -0.911 | . | . | -0.661 |
| Hungary | 0.289 | -0.862 | 1.783 | -0.639 | 0.487 | -0.660 |
| Iceland | 0.984 | 1.347 | 0.885 | . | -0.769 | -0.647 |
| India | -0.458 | -0.017 | 0.346 | 0.808 | 0.194 | -0.268 |
| Indonesia | 0.288 | 0.862 | -1.181 | 0.855 | 0.655 | 1.802 |
| Iran | -1.638 | 0.896 | 0.032 | -0.079 | -1.355 | 2.242 |
| Iraq | -1.837 | 1.741 | -0.103 | . | -0.894 | 2.125 |
| Ireland | 0.817 | 0.908 | 1.334 | -1.480 | -0.936 | -0.644 |
| Italy | 0.971 | 0.090 | 1.603 | -0.452 | 0.613 | -0.627 |
| Jamaica | 0.498 | . | -0.058 | -0.686 | . | -0.660 |
| Japan | 0.725 | 0.199 | 0.256 | -0.266 | 1.743 | -0.657 |
| Jordan | -0.653 | 0.012 | -0.462 | . | -1.271 | 2.184 |
| Kenya | -0.080 | . | -0.687 | . | . | -0.355 |
| Korea, South | 0.671 | -0.709 | -1.001 | 0.015 | 2.245 | -0.657 |
| Kuwait | -1.684 | . | -0.103 | . | . | 2.036 |
| Kyrgyzstan | -0.367 | -0.677 | . | . | 0.822 | 1.598 |
| Latvia | 1.271 | -0.176 | 1.334 | -0.732 | 0.948 | -0.661 |
| Lebanon | -0.658 | . | -0.103 | . | . | 0.922 |
| Libya | -1.106 | . | -0.103 | . | . | 2.182 |
| Lithuania | 1.390 | -0.653 | 0.885 | -0.826 | 1.492 | -0.659 |
| Luxembourg | 0.567 | 0.382 | 0.885 | -0.919 | 0.738 | -0.603 |
| Malawi | 0.676 | . | -0.462 | . | . | -0.233 |
| Malaysia | -0.866 | . | -0.642 | 2.069 | -0.224 | 1.128 |
| Mali | 0.048 | . | . | . | -1.103 | 2.096 |
| Malta | 0.683 | -1.585 | 0.840 | -0.172 | 0.027 | -0.632 |
| Mexico | -0.168 | -0.170 | -0.462 | 0.995 | -0.936 | -0.661 |
| Moldova | 0.618 | -0.311 | . | . | 1.031 | -0.556 |
| Montenegro | 0.268 | . | . | . | 1.199 | -0.101 |
| Morocco | -0.939 | -0.395 | 0.256 | 0.482 | -1.355 | 2.240 |
| Mozambique | -0.521 | . | -1.136 | . | . | 0.053 |
| Namibia | 0.891 | . | -0.462 | . | . | -0.647 |
| Nepal | 0.654 | . | -0.462 | . | . | -0.533 |
| Netherlands | 0.308 | 1.964 | 1.783 | -1.013 | 0.864 | -0.491 |
| New Zealand | -0.556 | 1.831 | 1.738 | -1.760 | -0.559 | -0.629 |
| Nigeria | -0.443 | -0.500 | -0.911 | . | -1.396 | 0.829 |
| North Macedonia | 0.467 | -0.619 | . | . | 0.655 | 0.342 |
| Norway | 0.661 | 1.286 | 1.289 | -1.339 | -0.475 | -0.602 |
| Pakistan | -1.544 | -0.834 | -1.181 | -0.219 | 0.152 | 2.145 |
| Panama | 0.433 | . | -1.315 | 1.649 | . | -0.652 |
| Peru | 0.801 | -1.019 | -1.091 | 0.201 | -0.894 | -0.661 |
| Philippines | -0.304 | -1.888 | -0.373 | 1.602 | -0.810 | -0.499 |
| Poland | 1.002 | 0.279 | 0.885 | 0.388 | -0.350 | -0.661 |
| Portugal | 1.289 | -0.559 | -0.597 | 0.155 | -0.769 | -0.636 |
| Romania | 0.860 | -0.954 | -0.462 | 1.415 | 0.236 | -0.650 |
| Russia | -1.734 | 0.112 | -0.058 | 1.556 | 1.450 | -0.322 |
| Rwanda | -2.164 | . | . | . | -1.187 | -0.526 |
| Saudi Arabia | -1.139 | 0.807 | -0.103 | . | -0.434 | 2.090 |
| Senegal | 0.215 | . | -0.687 | . | . | 2.092 |
| Serbia | 0.811 | -0.531 | -0.687 | 1.229 | 0.236 | -0.570 |
| Sierra Leone | 0.126 | . | -0.911 | . | . | 0.961 |
| Singapore | -0.325 | -0.601 | -0.911 | 0.668 | 1.073 | -0.244 |
| Slovakia | 1.258 | -0.954 | 0.525 | 2.069 | 1.283 | -0.661 |
| Slovenia | 1.244 | -0.241 | -0.597 | 0.528 | 0.111 | -0.573 |
| South Africa | 1.209 | -0.491 | 1.109 | . | -0.517 | -0.612 |
| Spain | 1.289 | 0.418 | 0.481 | -0.125 | 0.069 | -0.572 |
| Sri Lanka | -0.736 | . | -0.238 | . | . | -0.380 |
| Suriname | 0.849 | . | 0.301 | 1.182 | . | -0.086 |
| Sweden | 1.054 | 2.928 | 1.379 | -1.339 | 0.278 | -0.485 |
| Switzerland | 1.158 | 0.814 | 1.244 | -1.199 | 1.157 | -0.541 |
| Syria | -1.816 | . | -0.238 | . | . | 1.984 |
| Taiwan | 0.724 | 0.167 | -1.046 | -0.079 | 1.952 | -0.648 |
| Tanzania | 0.084 | -0.186 | -0.597 | . | -0.517 | 0.395 |
| Thailand | -1.342 | . | -0.911 | 0.201 | -0.601 | -0.368 |
| Trinidad and Tobago | 0.960 | . | -1.091 | -0.592 | -1.396 | -0.514 |
| Turkiye | -1.366 | -1.066 | -0.148 | 0.295 | -0.015 | 2.230 |
| Uganda | -1.539 | -1.690 | . | . | -0.936 | -0.309 |
| Ukraine | -0.641 | -0.175 | -0.687 | . | 1.659 | -0.632 |
| United Arab Emirates | -1.533 | . | -0.103 | . | . | 1.318 |
| United Kingdom | -0.439 | 1.137 | 2.187 | -1.153 | 0.194 | -0.534 |
| United States | 0.055 | 0.841 | 2.277 | -0.919 | -0.852 | -0.635 |
| Uruguay | 0.683 | -0.098 | -0.193 | 0.061 | -0.852 | -0.660 |
| Venezuela | -1.096 | -0.628 | -1.271 | 0.995 | -1.271 | -0.658 |
| Vietnam | -1.544 | 0.756 | -0.911 | 0.482 | 0.445 | -0.658 |
| Zambia | -0.020 | . | -0.597 | . | -0.685 | -0.645 |
| Zimbabwe | -1.564 | -0.376 | . | . | -1.313 | -0.630 |

Table OA.3: Robust regression

| | (1)<br>CC | (2)<br>CC | (3)<br>CC | (4)<br>CC | (5)<br>CC |
|---|---|---|---|---|---|
| Morality | 0.034 | | | | |
| | (0.126) | | | | |
| Individ. | | 0.218* | | | |
| | | (0.105) | | | |
| Power dist. | | | -0.235* | | |
| | | | (0.114) | | |
| Long-term | | | | 0.255* | |
| | | | | (0.111) | |
| Muslim | | | | | -0.562*** |
| | | | | | (0.080) |
| Controls | No | No | No | No | No |
| Estimator | Robust | Robust | Robust | Robust | Robust |
| Obs. | 79 | 101 | 67 | 89 | 115 |
| R-sq. | 0.00 | 0.04 | 0.06 | 0.06 | 0.30 |

Note: Robust regression coefficient estimates (see Hamilton 1991) with robust standard errors in parentheses. Dependent variable: constitutional compliance. Constant omitted. + p<.1, * p<.05, ** p<.01, *** p<.001.

Table OA.4: Effect of religious adherence on compliance

| | (1) CC | (2) CC | (3) CC | (4) CC | (5) CC |
|---|---|---|---|---|---|
| Catholic share | 1.173*** | | | 0.690+ | 0.738 |
| | (0.295) | | | (0.359) | (0.633) |
| Protestant share | | 1.268*** | | 0.732+ | 0.750 |
| | | (0.349) | | (0.393) | (0.693) |
| Muslim share | | | -1.580*** | -1.205*** | -1.177* |
| | | | (0.244) | (0.316) | (0.522) |
| Orthodox share | | | | | 0.288 |
| | | | | | (0.597) |
| Buddhist share | | | | | -0.599 |
| | | | | | (0.787) |
| Hindu share | | | | | 0.435 |
| | | | | | (0.692) |
| Controls | Yes | Yes | Yes | Yes | Yes |
| Estimator | OLS | OLS | OLS | OLS | OLS |
| Obs. | 115 | 115 | 115 | 115 | 115 |
| R-sq. | 0.19 | 0.15 | 0.32 | 0.36 | 0.37 |

Note: OLS regression coefficient estimates with robust standard errors in parentheses. Dependent variable: constitutional compliance. Independent variables are (non-standardized) religion population shares. Constant omitted. + p<.1, * p<.05, ** p<.01, *** p<.001.

Table OA.5: Effect of religious majority on compliance

| | (1) CC | (2) CC | (3) CC | (4) CC | (5) CC |
|---|---|---|---|---|---|
| Catholic majority | 0.417+ | | | 0.275 | 0.277 |
| | (0.248) | | | (0.261) | (0.288) |
| Protestant majority | | 0.706*** | | 0.628** | 0.617* |
| | | (0.183) | | (0.206) | (0.243) |
| Muslim majority | | | -1.121*** | -1.036*** | -1.027*** |
| | | | (0.216) | (0.222) | (0.245) |
| Orthodox majority | | | | | 0.090 |
| | | | | | (0.325) |
| Buddhist majority | | | | | -0.350 |
| | | | | | (0.495) |
| Hindu majority | | | | | 0.182 |
| | | | | | (0.510) |
| Controls | Yes | Yes | Yes | Yes | Yes |
| Estimator | OLS | OLS | OLS | OLS | OLS |
| Obs. | 115 | 115 | 115 | 115 | 115 |
| R-sq. | 0.12 | 0.12 | 0.27 | 0.29 | 0.30 |

Note: OLS regression coefficient estimates with robust standard errors in parentheses. Dependent variable: constitutional compliance. Independent variables are binary indicators for religious majorities. Constant omitted. + p<.1, * p<.05, ** p<.01, *** p<.001.

Table OA.6: Effect independent of democracy, Lewbel estimator

| | (1)<br>CC | (2)<br>CC | (3)<br>CC | (4)<br>CC | (5)<br>CC |
|---|---|---|---|---|---|
| Morality | -0.078 | | | | |
| | (0.091) | | | | |
| Individ. | | 0.279* | | | |
| | | (0.129) | | | |
| Power dist. | | | -0.086 | | |
| | | | (0.099) | | |
| Long-term | | | | -0.008 | |
| | | | | (0.167) | |
| Muslim | | | | | -0.370*** |
| | | | | | (0.097) |
| Democracy | 1.615*** | 1.290*** | 1.752*** | 1.498*** | 1.147*** |
| | (0.255) | (0.225) | (0.331) | (0.231) | (0.212) |
| Controls | Yes | Yes | Yes | Yes | Yes |
| Estimator | Lewbel | Lewbel | Lewbel | Lewbel | Lewbel |
| Obs. | 68 | 98 | 63 | 86 | 115 |
| R-sq. | 0.49 | 0.47 | 0.52 | 0.43 | 0.53 |
| F-stat | 11.06 | 6.18 | 21.13 | 5.02 | 94.71 |
| Stock-Yogo | 6.61 | 6.61 | 6.65 | 6.56 | 6.61 |
| Robust CI | [-.38],[-.09] | [.05],[.49] | [-.20],[.07] | [-.55],[.50] | [-.41],[-.27] |
| J-stat | 12.02 | 10.42 | 13.73 | 12.62 | 9.69 |

Note: 2SLS regression coefficient estimates with robust standard errors in parentheses. Dependent variable: constitutional compliance. External instruments are supplemented with heteroskedasticity-based instrumental variables constructed using Lewbel's (2012) method (see Baum and Schaffer 2026). "F-stat" refers to the Kleibergen-Paap rk Wald F statistic. Stock and Yogo (2005)-critical values for weak identification are reported for 20% maximal IV relative bias. Instruments are considered weak if the F-test falls below the "Stock-Yogo" critical value. "Robust CI" refers to an identification-robust confidence set of the independent (culture) variable of interest that is robust to weak instruments (see Andrews 2018; Sun 2018). The weak-IV robust estimate is significant at the 5%-level if the confidence set does not enclose zero. "J-stat" refers to the Sargan-Hansen test of overidentifying restrictions. A * on J-stat indicates that the null hypothesis of instrument validity can be rejected at the 5%-level. Constant omitted. + p<.1, * p<.05, ** p<.01, *** p<.001.

Table OA.7: Effect of individualism & Muslim share on dimensions

| | (1) CC | (2) Prop | (3) Polit | (4) Civil | (5) Basic |
|---|---|---|---|---|---|
| Individ. | 0.388* | 0.465** | 0.259 | 0.280 | 0.336* |
| | (0.165) | (0.144) | (0.162) | (0.174) | (0.163) |
| Muslim | -0.559*** | -0.472*** | -0.521*** | -0.664*** | -0.297* |
| | (0.109) | (0.102) | (0.101) | (0.115) | (0.143) |
| Controls | Yes | Yes | Yes | Yes | Yes |
| Estimator | Lewbel | Lewbel | Lewbel | Lewbel | Lewbel |
| Obs. | 98 | 98 | 97 | 98 | 98 |
| R-sq. | 0.37 | 0.39 | 0.37 | 0.41 | 0.13 |
| F-stat | 6.75 | 6.75 | 6.68 | 6.75 | 6.75 |
| Stock-Yogo | 6.14 | 6.14 | 6.14 | 6.14 | 6.14 |
| Rob. CI (IDV) | [.39],[1.04] | [.57],[1.09] | [.16],[1.35] | [.15],[.79] | [.34],[1.08] |
| Rob. CI (ISL) | [-.72],[-.32] | [-.61],[-.26] | [-.75],[-.29] | [-.93],[-.40] | [-.49],[-.01] |

Note: 2SLS regression coefficient estimates with robust standard errors in parentheses. Dependent variables: constitutional compliance and its dimensions. External instruments are supplemented with heteroskedasticity-based instrumental variables constructed using Lewbel's (2012) method (see Baum and Schaffer 2026). "F-stat" refers to the Kleibergen-Paap rk Wald F statistic. Stock and Yogo (2005)-critical values for weak identification are reported for 20% maximal IV relative bias. Instruments are considered weak if the F-test falls below the "Stock-Yogo" critical value. "Rob. CI" refers to an identification-robust confidence set of one of the independent (culture) variables of interest that is robust to weak instruments (see Andrews 2018; Sun 2018). The weak-IV robust estimate is significant at the 5%-level if the confidence set does not enclose zero. With two endogenous regressors, confidence sets are projection-based and, therefore, conservative. Constant omitted. + p<.1, * p<.05, ** p<.01, *** p<.001.

Table OA.8: Effect of individualism, alternative concepts

| | (1) CC | (2) CC | (3) CC |
|---|---|---|---|
| Kinship tightness | -0.324** | | |
| | (0.118) | | |
| Tightness (32) | | -0.115 | |
| | | (0.137) | |
| Tightness (57) | | | -0.309+ |
| | | | (0.181) |
| Controls | Yes | Yes | Yes |
| Estimator | OLS | OLS | OLS |
| Obs. | 98 | 30 | 52 |
| R-sq. | 0.19 | 0.37 | 0.29 |

Note: OLS regression coefficient estimates with robust standard errors in parentheses. Dependent variable: constitutional compliance. Independent variables are kinship tightness (Enke 2019) and tightness/looseness (Gelfand et al. 2011; Eriksson et al. 2021). Constant omitted. + p<.1, * p<.05, ** p<.01, *** p<.001.

Table OA.9: Causal mediation analysis with parallel mediators

| | (1) | (2) | (3) | (4) | (5) |
|---|---|---|---|---|---|
| Natural direct effect | | | | | |
| Morality | -0.145 | | | | |
| | (0.090) | | | | |
| Individ. | | -0.125+ | | | |
| | | (0.067) | | | |
| Power dist. | | | 0.117 | | |
| | | | (0.094) | | |
| Long-term | | | | 0.204*** | |
| | | | | (0.051) | |
| Muslim | | | | | -0.173** |
| | | | | | (0.067) |
| NIE (Veto players) | | | | | |
| Morality | 0.103+ | | | | |
| | (0.053) | | | | |
| Individ. | | 0.196** | | | |
| | | (0.063) | | | |
| Power dist. | | | -0.183 | | |
| | | | (0.129) | | |
| Long-term | | | | -0.002 | |
| | | | | (0.020) | |
| Muslim | | | | | -0.077+ |
| | | | | | (0.043) |
| NIE (Civil society) | | | | | |
| Morality | 0.057 | | | | |
| | (0.062) | | | | |
| Individ. | | 0.150** | | | |
| | | (0.046) | | | |
| Power dist. | | | -0.188* | | |
| | | | (0.081) | | |
| Long-term | | | | 0.025 | |
| | | | | (0.076) | |
| Muslim | | | | | -0.269*** |
| | | | | | (0.065) |
| Total effect | | | | | |
| Morality | 0.015 | | | | |
| | (0.107) | | | | |
| Individ. | | 0.221** | | | |
| | | (0.080) | | | |
| Power dist. | | | -0.254* | | |
| | | | (0.104) | | |
| Long-term | | | | 0.227* | |
| | | | | (0.109) | |
| Muslim | | | | | -0.518*** |
| | | | | | (0.069) |
| Obs. | 79 | 101 | 67 | 89 | 115 |

Note: Decomposed effect estimates with robust standard errors in parentheses. Dependent variable: constitutional compliance. No control variables are included. Effects refer to a one-unit (one SD) increase in the respective cultural trait. + p<.1, * p<.05, ** p<.01, *** p<.001.

## Additional References